\documentclass[twocolumn]{IEEEtran}
\usepackage{cite}

\ifCLASSINFOpdf
   \usepackage[pdftex]{graphicx}
  \DeclareGraphicsExtensions{.pdf,.jpeg,.png}
\else
  \usepackage[dvips]{graphicx}
  \DeclareGraphicsExtensions{.eps}
\fi
\usepackage{relsize}
\usepackage[cmex10]{amsmath}
\usepackage{algorithmic}

\usepackage{array}
\usepackage{fixltx2e}
\usepackage{dblfloatfix}

\ifCLASSOPTIONcaptionsoff
  \usepackage[nomarkers]{endfloat}
 \let\MYoriglatexcaption\caption
 \renewcommand{\caption}[2][\relax]{\MYoriglatexcaption[#2]{#2}}
\fi
\usepackage{url}

\usepackage{epsfig}
\usepackage{multicol}
\usepackage{verbatim}
\usepackage{amssymb}
\usepackage{epstopdf}
\usepackage{amsthm}
\usepackage{soul}
\usepackage{amsfonts}%
\usepackage{enumerate}%
\usepackage{psfrag}
\usepackage{setspace}
\usepackage{dsfont}
\usepackage{balance}

\usepackage{algorithm}
\usepackage{float}

\usepackage{mdwmath}
\usepackage{mdwtab}
\usepackage[section]{placeins}
\usepackage{caption}
\usepackage{subcaption}
\usepackage{xcolor}
\usepackage{mathrsfs}

\newtheorem{theorem}{Theorem}
\newtheorem{proposition}{Proposition}
\newtheorem{remark}{Remark}
\newtheorem{definition}{Definition}

\renewcommand{\vec}[1]{\mathbf{#1}}

\newenvironment{rcases}
  {\left.\begin{aligned}}
  {\end{aligned}\right\rbrace}

\begin{document}

\title{Secure Degrees of Freedom of the Gaussian MIMO Wiretap and MIMO Broadcast Channels with Unknown Eavesdroppers}


\author{\IEEEauthorblockN{Mohamed Amir and Tamer Khattab}\\
\IEEEauthorblockA{Electrical Engineering, Qatar University\\
Email: mohamed.amir@qu.edu.qa, tkhattab@ieee.org}
\thanks{This research was made possible by NPRP 5-559-2-227 grant from the Qatar National Research Fund (a member of The Qatar Foundation). The statements made herein are solely the responsibility of the authors.}}
\maketitle

\begin{abstract}
We investigate the secure degrees of freedom (SDoF) of the wiretap and the $K$ user Gaussian broadcast  channels with multiple antennas at the transmitter, the legitimate receivers and an unknown number of eavesdroppers each with a number of antennas less than or equal to a known value $N_E$. The channel matrices between the legitimate transmitter and the receivers are available everywhere, while the legitimate pair have no information about the eavesdroppers' channels. We provide the exact sum SDoF for the considered system. A new comprehensive upperbound is deduced and a new achievable scheme based on utilizing jamming is exploited. We prove that cooperative jamming is SDoF optimal even without the eavesdropper CSI available at the transmitters.   

\end{abstract}

\vspace{-4mm}
\section{Introduction}
The noisy wiretap channel was first studied by Wyner \cite{wyner}, in which a
legitimate transmitter (Alice) wishes to send a message to a legitimate receiver (Bob), and hide it from an eavesdropper (Eve). Wyner proved that Alice can send positive secured rate using channel coding. He derived capacity-equivocation region for the degraded
wiretap channel, then the generalization to the
general wiretap channel was done by Csiszar and Korner\cite{csi}.  Leung-Yan-
Cheong and Hellman \cite{leu} then extended the results to the Gaussian wiretap channel case.\\

Substantial work was done hereafter to study the information theoretic physical layer security for different network models. The relay assisted wiretap channel was studied in \cite{secop}. The degrees of freedom (DoF) region of multiple access channel was presented in \cite{sennur_MAC}.  
Using MIMO systems for securing the message was an intuitive extension due to the spatial gain provided by multiple antennas.
The MIMO wiretap channel was studied in~\cite{mimo_wire, mimo_secure,mimo_note,mimo_hass,yener_coop,yener_mimo,yener_bc,mimo_shafie,mimo_confidential} and the secrecy capacity was identified in \cite{mimo_hass}.  
All these previous works assumed the availability of either partial or complete channel state information (CSI). Given that the eavesdropper is passive, it is more practical to assume that the eavesdropper CSI is completely unknown. Papers~\cite{yener_mimo,yener_bc} study the secrecy capacity and secure DoF for different MIMO channels when the eavesdropper channel is arbitrarily varying and its channel states are known to the eavesdropper only.

Meanwhile, the idea of cooperative jamming was proposed in \cite{yener_coop},
 where some of the $K$--single antenna transmitters jam the eavesdropper by transmitting independent and identically distributed (i.i.d.)
Gaussian noise to improve the sum secrecy rate. Cooperative
jamming was then used for deriving the SDoF for different networks. In \cite{sennur_MAC}, cooperative jamming was used to zero the DoF decoded by the eavesdropper and prove that the MAC channel with single antenna nodes has $\frac{K(K-1)}{K(K-1)+1}$ DoF.\\ 

Inspired by cooperative jamming, we devise a technique called \textit{partial jamming}, where some of the transmitter's DoF are used to send jamming signals, while the remaining DoF are used for secure signal transmission.  We utilize partial jamming to investigate the MIMO wiretap channel, the $2$--user MIMO broadcast channel and $K$--user MIMO \textit{symmetric}\footnote{The term symmetric is used to indicate the case of equal number of antennas at all receivers} broadcast channel with unknown eavesdroppers' CSI at the transmitter and receivers, under fading eavesdroppers' channels. We provide a new upperbound for the achievable SDoF and determine the exact sum SDoF by providing an achievable scheme.  We show that our scheme is optimal and that the achievable bound and the new upperbound are tight. Compared to previous art, the novelty of this work can be summarized as follows:
\begin{itemize}
\item We prove that knowledge of eavesdropper's CSI does not increase SDoF for the presented channel models,
\item For the case of known eavesdropper channels with constant or time varying channels, we show that it has the same sum SDoF as the previous case,
\item We incorporate the more general scenario of multiple eavesdroppers, 
\item No restrictions are assumed on the relation between the number of antennas at the transmitter and the receivers, 
\item We address the general case where all the eigenvalues of the legitimate channel have non-zero values\footnote{The cases where some of the eigenvalues are equal to zero represent special degraded cases of the more general non-zero eigenvalues case, where the SDoF decreases for every zero eigenvalue till it collapses to the trivial case of zero SDoF for all-zero eigenvalues.}, 
\item For the special case of a single eavesdropper, our proposed scheme achieves a sum SDoF superior to those reported in the literature. 
\end{itemize}
 
The paper is organized as follows. Section \ref{model} defines the system model and the secrecy constraints. The main results are presented in Section \ref{results}. In Section \ref{bound}, the new upperbound is derived and the achievable scheme is presented in Section \ref{scheme}.  The paper is concluded in Section \ref{con}. We use the following notation, $\vec{a}$ for vectors, $\vec{A}$ for matrices, $\vec{A}^{\dagger}$ for the hermitian transpose of $\vec{A}$, $[A]^+$ for the $\max{A,0}$ and $\text{Null}(\vec{A})$ to define the nullspace of $\vec{A}$. 
\vspace{-3mm}

\section{System model}\label{model}

\begin{figure}
  \begin{center}
\hspace{-4mm} \includegraphics[width=.4\textwidth]{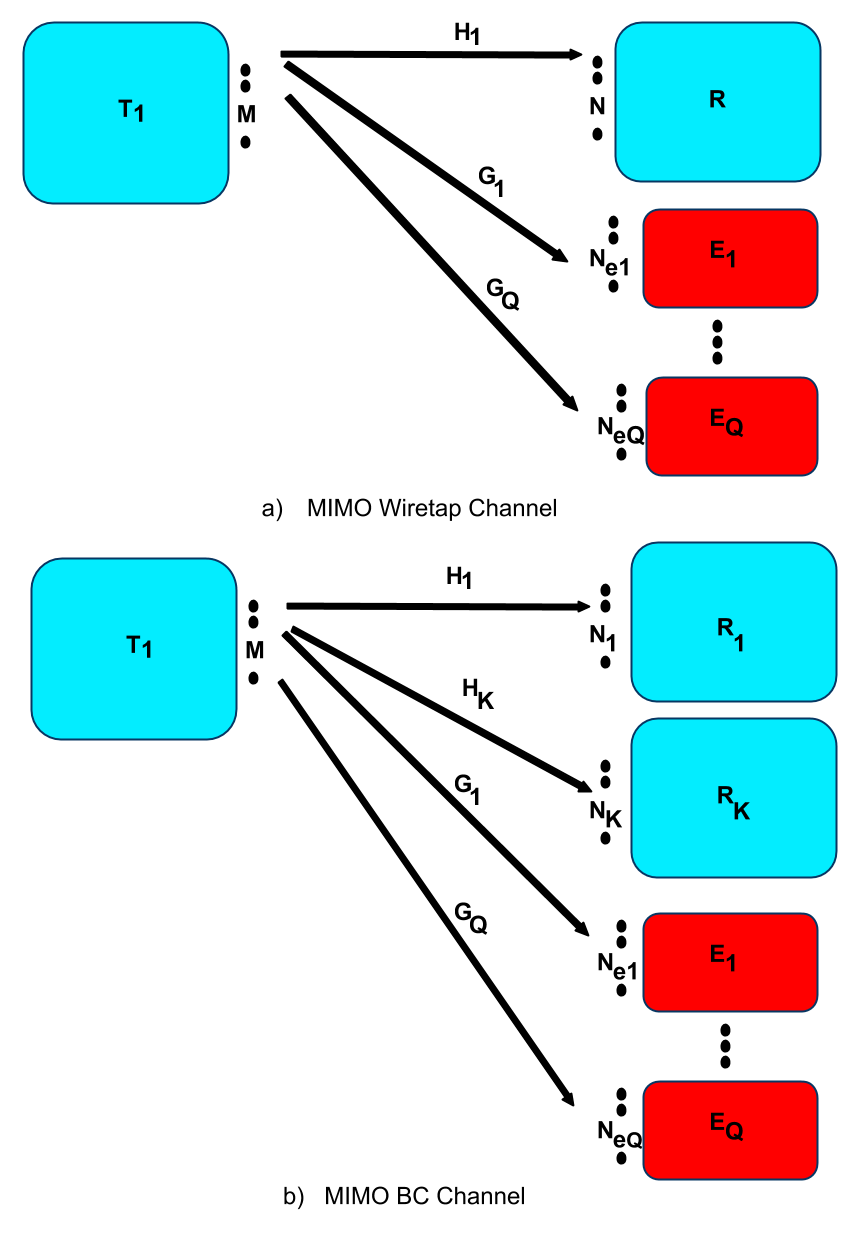}\vspace{-5mm}
\caption{\small System model}
  \label{sys}
  \end{center} 
  \end{figure}

We consider two communication systems, the MIMO wiretap channel (Fig.~\ref{sys}-a) and the MIMO broadcast channel with a single transmitter and $K$ receivers (Fig.~\ref{sys}-b), all with the existence of unknown number, $Q$, of  passive eavesdroppers. In both systems, the transmitter is equipped with $M$ antennas while receiver $i$ is equipped with $N_i$ antennas. The $j$th eavesdropper is equipped with $N_{Ej}\leq N_E$ antennas. 

Let $\vec{x}$ denote the $M \times 1$ vector of Gaussian i.i.d coded symbols to be transmitted by the transmitter. We can
write the received signal at the $i$th legitimate receiver at time (sample) $l$ as
\begin{equation}\label{Received_signal}
\vec{y}_i(l)=\vec{H}_i \vec{V}\vec{x}(l)+ \vec{n}_i(l)
\end{equation}
\noindent where $\vec{H}_i$ is the $N_i \times M$ matrix containing
the channel coefficients independently drawn from a continuous distribution from the transmitter to the receiver $i$, $i\in{1:K}$ for the BC channel and $i=1$ for the MIMO wiretap channel.
The received signal at the $j$th eavesdropper is
\begin{equation}\label{Received_signal}
\vec{z}_{j}(l)=\vec{G}_{j}(l) \vec{V}\vec{x}(l)+ \vec{n}_{Ej}(l),
\end{equation}

\noindent where  $\vec{G}_{j}(l)$ is the $N_{Ej} \times M$ matrix containing the the i.i.d time varying channel coefficients from transmitter $i$ to the eavesdropper $j$ drawn from a continuous distribution with mean $\eta$ and variance $\sigma_e^2$,  $\vec{V}$ is the precoding unitary matrix (i.e. $\vec{V}\vec{V}^\dagger = \vec{I}$) at the transmitter, $\vec{n}(l)$ and $\vec{n}_{Ej}(l)$ are the $N\times 1$  and the $N_{Ej}\times 1$ additive white Gaussian
noise vectors with zero mean and variance $\sigma^2$ at the legitimate receiver and the $j$th eavesdropper, respectively. It is assumed that the maximum number of antennas any eavesdropper can possess; namely, $N_E$, is known to the transmitter, while the eavesdroppers' channels, $\vec{G}_{j}(l)$, are unknown. We focus on the case $N_E< M$ to avoid the trivial zero SDoF case.  

We define the $M \times 1$ channel input from the legitimate transmitter as
\begin{equation}
\vec{\bar{x}}(l)= \vec{V} \vec{x}(l).
\end{equation}

	\normalsize
    \vspace{1.5mm}
The transmitter intends to send a message $W_i$ to legitimate receiver $i$ over $n$ channel uses (samples) simultaneously while preventing the eavesdroppers from decoding the message.  The encoding occurs under  a constrained power given by
\begin{equation}
 \text{E}\left\{\vec{\bar{x}}\vec{\bar{x}}^{\dagger}\right\} \leq P
\end{equation}

Expanding the notations over $n$ channel extensions we have
$$\vec{H}_i^n = \vec{H}_i(1), \vec{H}_i(2), \ldots, \vec{H}_i(n),$$ 
$$\vec{G}_{j}^{n} = \vec{G}_{j}(1), \vec{G}_{j}(2), \ldots, \vec{G}_{j}(n)$$ 
and similarly the time extended channel input, $\vec{X}^n$, time extended channel output at legitimate receiver $i$,  $\vec{Y}_i^n$ and time extended channel output at eavesdropper $j$, $\vec{Z}_{j}^n$ as well as noise at legitimate receiver, $\vec{n}^n$ and noise at eavesdroppers, $\vec{n}_{Ej}^n$.

At each transmitter, the message $W_i$ is uniformly and independently chosen from a set of possible secret messages for  receiver $i$, $\mathcal{W}_i = \{1,2, \ldots, 2^{nR_i}\}$.  The rate for message $W_i$ is $R_i \triangleq \frac{1}{n} \log\left|\mathcal{W}_i\right|$, where $|\cdot|$ denotes the cardinality of the set. The transmitter uses a stochastic encoding function $f: W_1, W_2, \ldots, W_K  \longrightarrow \vec{X}^n$ to map the secret messages into a transmitted symbol.  The receiver $i$ has a decoding function $\phi: \vec{Y}_i^n \longrightarrow \hat{W}_i$, where $\hat{W}_i$ is an estimate of $W_i$.
\begin{definition}
A secure rate tuple $(R_1, R_2, \ldots, R_K)$ is said to be achievable if for any $\epsilon > 0$ there exist $n$-length codes such that the legitimate receiver can decode the messages reliably, i.e.,
\begin{equation}
\text{Pr}\{(W_1,W_2, \ldots, W_K) \neq (\hat{W}_1, \hat{W}_2\, \ldots, \hat{W}_K)\} \leq \epsilon
\end{equation}
and the messages are kept information-theoretically secure against the eavesdroppers, i.e.,
\begin{equation}\label{eqn:cond}
H(W_1, W_2\ldots, W_K|\vec{Z}_{j}^{n})\geq H(\hat{W}_1, \hat{W}_2,\ldots, \hat{W}_K)-\epsilon \\
\end{equation}
\begin{equation}
H(W_i|\vec{Z}_{j}^{n})\geq H(\hat{W}_i) - \epsilon \;\;\; \forall \; i \in \{1,2,..,K\}
\end{equation}
\noindent where $H(\cdot)$ is the Entropy function and~\eqref{eqn:cond} implies the secrecy for any subset $\mathcal{S} \subset \{1,2,..,K\}$ of messages including individual messages~\cite{sennur_MAC}.
\end{definition}
\vspace{-1mm}
\begin{definition}
The sum SDoF is defined as
\begin{equation}
D_s = \lim_{P\rightarrow \infty} \sup{\sum_i \frac{R_i}{\frac{1}{2}\log P}},
\end{equation}
\noindent where the supremum is over all achievable secrecy rate tuples $(R_1, R_2, \ldots, R_K)$, $D_s = d_1 + d_2+\ldots +d_K$, and $d_i$ is the SDoF of receiver $i$. 
\end{definition}
\vspace{-6mm}
\section{Main results}\label{results}
\vspace{-1mm}
\begin{theorem}
The SDoF of the MIMO wiretap channel is
\begin{equation}
d^w =
\begin{cases}
    M-N_E ,& \text{for  } M \leq N_1+N_E \\
    N_1,              & \text{for  } M > N_1+N_E 
\end{cases},
\end{equation}

the SDoF region of the two user BC channel is
\begin{equation}
\begin{rcases}
d^{bc}_1 & \leq \min (N_1, M-N_E ) \\
d^{bc}_2 & \leq \min (N_2, M-N_E ) \\
\nonumber d^{bc}_1+d^{bc}_2& \leq \min (N_1+N_2, M-N_E ) 
\end{rcases} 
\text{if } M \geq N_1 \geq N_2
\end{equation}

\begin{equation}\label{r1}
\begin{rcases}
d^{bc}_2 &\leq \min (N_2, M-N_E) \\ 
\nonumber d^{bc}_1+d^{bc}_2 &\leq  M-N_E 
\end{rcases} 
\nonumber\text{if } N_2 < M \leq N_1
\end{equation}
\begin{equation}
d^{bc}_1+d^{bc}_2 \leq M-N_E
\;\;\;\;\;\; \text{  if }  M < N_2 \leq N_1
\end{equation}

and the sum SDoF of the $K$ user symmetric BC channel is

\begin{equation}
D^{bc} = \sum_{i=1}^K d^{bc}_i =
   \min (M-N_E, KN)  
\end{equation}
\end{theorem}
\begin{proof}
The converse proof for this theorem is provided in Section~\ref{bound}, while the achievability is provided in Section~\ref{scheme}.
\end{proof}
 
\section{Converse}\label{bound}

We will first derive the upperbound for the broadcast channel case and then derive the MIMO wiretap channel upperbound as a special case achieved by setting $Y_i$ to null for $i\in \{2,..,K\}$. 
\begin{eqnarray}
\nonumber n \sum_{i=1}^K R_i\hspace{-5mm}&&\leq I(W_1,W_2,\ldots, W_K;\vec{Y}_1^n,\vec{Y}_2^n,\ldots, \vec{Y}_K^n)\\
\nonumber&&- I(W_1,W_2,\ldots, W_K;\vec{Z}^n)\\
\nonumber&&\leq I(W_1,W_2, \ldots, W_K;\vec{Y}_1^n,\vec{Y}_2^n,\ldots, \vec{Y}_K^n, \vec{Z}^n)\\
\nonumber&&- I(W_1,W_2,\ldots, W_K; \vec{Z}^n)\\
\nonumber&&= I(W_1,W_2, \ldots, W_K;\vec{Y}_1^n,\vec{Y}_2^n, \ldots, \vec{Y}_K^n|\vec{Z}^n)\\
\nonumber&&\leq I(\vec{X}^n;\vec{Y}_1^n, \vec{Y}_2^n, \ldots, \vec{Y}_K^n|\vec{Z}^n)\\
\nonumber&&=h(\vec{Y}_1^n,\vec{Y}_2^n, \ldots, \vec{Y}_K^n|\vec{Z}^n)\\
\nonumber&&-h(\vec{Y}_1^n,\vec{Y}_2^n, \ldots, \vec{Y}_K^n|\vec{Z}^n, \vec{X}^n)\\
\nonumber&&=h(\vec{Y}_1^n,\vec{Y}_2^n, \ldots, \vec{Y}_K^n|\vec{Z}^n)-h(\vec{n}^n_1,\vec{n}^n_2, \ldots, \vec{n}^n_K)\\
\nonumber&&=h(\vec{Y}_1^n,\vec{Y}_2^n, \ldots, \vec{Y}_K^n,\vec{Z}^n)-h(\vec{Z}^n)+C_1\\
\nonumber&&\leq h(\vec{X}^n,\vec{Y}_1^n,\vec{Y}_2^n, \ldots, \vec{Y}_K^n,\vec{Z}^n)-h(\vec{Z}^n)+C_1\\
\nonumber&&= h(\vec{X}^n)+h(\vec{Y}_1^n,\vec{Y}_2^n, \ldots, \vec{Y}_K^n,\vec{Z}^n|\vec{X}^n)-h(\vec{Z}^n)\\
\nonumber&&+C_1\\ 
\nonumber&&= h(\vec{X}^n)-h(\vec{Z}^n)+C_2\\ 
&&\leq \sum_{m=1}^{M}h({\vec{x}}_{m}^n)-h(\vec{Z}^n)+C_2 \label{up}
\end{eqnarray}

where $\vec{x}_{m}^n$ is the $m$th row of $\vec{X}^n$ and $x_m$ is the $m$th value of $\vec{x}$. Let $\vec{B}$ be a permutation matrix when multiplied by $\vec{z}$ results in a vector $\vec{\bar{z}}$ with the $m$th element ($m \in \{1, 2, \ldots, M\}$) depending on ${x}_m$ and  $e_m$, where $e_{m}$ are constants depending on $\vec{B}$ and $G_{i \; : \; i\in \{1,2,\ldots,K\}}$.
\begin{eqnarray}
\vec{\bar{z}}=\vec{B}\vec{z}=
\begin{bmatrix}
e_1 x_1 \\
e_2 x_2 \\
\vdots\\
e_{N_E-1} {x}_{N_E-1}\\
e_{N_E} {x}_{N_E}+..+e_{M} x_{M}
\end{bmatrix}
+ \vec{B} \vec{n}_E
\end{eqnarray}
and, 
\begin{eqnarray}
h(\bar{\vec{z}})=h(\vec{z})+\log|\vec{B}|  \label{perm}
\end{eqnarray}
Substituting (\ref{perm}) into (\ref{up}),

\begin{eqnarray}
\nonumber n\sum_{i=1}^K R_i & \leq & n\sum_{m=1}^{M}h({x}_{m})-nh(\bar{\vec{z}})\\
\nonumber & + & \log|\vec{B}|+C_2 \\
\nonumber \sum_{i=1}^K R_i & \leq & \sum_{m=1}^{M}h(x_m)-\sum^{N_E-1}_{m=1}h(e_m x_m)\\
\nonumber &-&h\left(\sum_{m=1}^{M-N_E+1}e_{m+N_E-1}x_{m+N_E-1}\right)\\
\nonumber &-& h(\vec{n}_E)+C_3\\ 
\nonumber & = & (M-N_E+1) \log P\\
\nonumber & - & \log \left(\left|\left|[e_{N_E} \ldots e_{M}]\right|\right|^2 P\right) +C_4
\end{eqnarray}

Consequently, the SDoF of the wiretap channel, which is calculated by setting $\{\vec{Y}_i = 0, i \in 2,...,K\}$ is upperbounded as

\begin{equation}\label{sec}
D^w \leq \text M-N_E
\end{equation}
\\
and the sum SDoF of the broadcast channel is upperbounded as
\begin{equation}\label{2}
D^{bc} \leq M-N_E.
\end{equation}
\\
Given the fact that any receiver $i$ has only $N_i$ antennas, the SDoF of the wiretap channel is upperbounded as 
\begin{equation}\label{sec}
D^w \leq \min(M-N_E,N_1),
\end{equation}
while the SDoF region of the broadcast channel is upperbounded as 
\begin{eqnarray}\label{2}
d^{\text{bc}}_i &\leq \min(M-N_E,N_i) \forall \text{ } i\in \{1,2,..,K\} \\
D^{\text{bc}} &\leq \min(M-N_E, \sum_{i=1}^K N_i)
\end{eqnarray}

\section{Achievable scheme}\label{scheme}

For securing the legitimate messages, the transmitter sends $N_E$ jamming signal vector $\vec{r}$ with random symbols.  Hence, the transmitted coded signal can be broken into legitimate signal, $\vec{s}$, and jamming signal, $\vec{r}$, such that 
$$\vec{x} = \left[\begin{array}{c} \vec{s}\\  \vec{r} \end{array}\right].$$
Accordingly, the precoder, $\vec{V}$ can be also broken into legitimate, $\vec{V}^L$, and jamming, $\vec{V}^J$ precoders such that
$$\vec{V} = \left[\begin{array}{cc} \vec{V}^L &  \vec{V}^J \end{array}\right].$$
  
Choosing $\vec{V}^J$ to be the unitary matrix, the jamming power becomes $P^J=\text{E}\{\text{tr}(\vec{x}^J\vec{x}^{J\dagger})\} = \alpha P$, where $\alpha$ is a constant controlled by the transmitter.

\begin{proposition}
The jamming signal, $\vec{x}^J$, overwhelms the signal space of the eavesdropper with the strongest channel, and the eavesdropper ends up decoding zero DoF of the legitimate messages. Accordingly, weaker eavesdroppers can decode zero DoF of the legitimate messages.\\
\end{proposition}

\begin{proof}
\begin{eqnarray}
\nonumber n R_e && \leq I(\vec{Z}^n; \vec{S}^n)\\
\nonumber  && = h(\vec{Z}^n) -h(\vec{Z}^n|\vec{S}^n)\\
\nonumber R_e && = h (\vec{Z}) - h (\vec{G}(\vec{V}^L\vec{s}+\vec{V}^J\vec{r}) + \vec{n}_E | \vec{s})\\
\nonumber && = h (\vec{Z}) - h (\vec{G} \vec{V}^J \vec{r}+ \vec{n}_E)\\
\nonumber && \leq\frac{1}{2}\log \frac{\left| \vec{I} \sigma^2+(\vec{G}\vec{V} \text{E}\{\vec{x}\vec{x}^{\dagger}\}\vec{V}^{\dagger}\vec{G}^{\dagger}) \right|}{\left| \vec{I} \sigma^2+(\vec{G}\vec{V}^J\text{E}\{\vec{r}\vec{r}^{\dagger}\}\vec{V}^{J\dagger}\vec{G}^{\dagger}) \right|}\\
&&\leq C,
\end{eqnarray}
\noindent where $C$ is a constant that does not depend on $P$ and known to the transmitter. Therefor, 
\begin{equation}
\nonumber \lim_{P\longrightarrow \infty} \frac{R_e(P)}{\frac{1}{2} \log_2 P} \leq \lim_{P\longrightarrow \infty} \frac{C}{\frac{1}{2} \log_2 P} = 0,
\end{equation}
\end{proof}

\begin{remark}\label{rem:1}
The constant eavesdropper rate comes from the fact that $P^J$ is controlled by the transmitter.  Hence, setting $P^J=\alpha P$ for some constant $\alpha$,  we guarantee a constant SNR at the eavesdropper and a constant rate independent of $P$. While the transmitter does not know the eavesdropper channel, it knows the maximum value of the effective channels created by the jamming $\vec{G}^\text{eff} = \frac{\vec{G}\vec{V}}{\vec{G}\vec{V}^J\vec{r}+\vec{I}_{N_E}\sigma}$. The value of $|\vec{G}^\text{eff}|^2$ is upperbounded by a constant $C^\text{eff}$, where
\begin{equation}
C^\text{eff}= \lim_{\substack{g^{k,l}\longrightarrow \infty,\\ k\in\{1,..,N_E\}, \\j\in\{1,..,M_i\}}}|\vec{G}^\text{eff}|^2,
\end{equation}
\noindent and $g^{k,l}$ is the element in the $k$th row and $l$th column of $\vec{G}$.
\end{remark}
\vspace{2mm}
The transmitter uses the rate difference to transmit perfectly secure messages using a stochastic encoder similar to the one described in~\cite{khan} according to the worst case scenario rate, $C$, for the strongest eavesdropper. Using results in~\cite{khan}, for some $\vec{U}_i$ projecting the signal at a jamming free space at receiver $i$ the achievable secrecy sum rate can be lowerbounded by

 \begin{eqnarray}
\nonumber \sum_{i=1}^K R_i \hspace{-3.5mm}& \geq \frac{1}{2} \log \left| \vec{I}+ \sum_{i=1}^K (\vec{U}_i\vec{H}_i\vec{V}\text{E}\{\vec{s}\vec{s}\}^{\dagger}\vec{V}^{\dagger}\vec{H}_i^{\dagger}\vec{U}_i^\dagger)\right| \hspace{-.7mm}-\hspace{-.7mm}R_e\\ 
 & \geq \hspace{-.5mm} \frac{1}{2} \log \left| \vec{I}+ \sum_{i=1}^K (\vec{U}_i\vec{H}_i\vec{V}\text{E}\{\vec{s}\vec{s}^{\dagger}\}\vec{V}^{\dagger}\vec{H}_i^{\dagger}\vec{U}_i^\dagger) \right| \hspace{-.7mm}-\hspace{-.7mm} C 
\label{secrate}
\end{eqnarray}

As $R_e$ is upperbounded by a constant for all values of $\vec{G}$ and $P$, a positive secrecy rate, which is monotonically increasing with $P$, is achieved. Computing the SDoF boils down to calculating the degrees of freedom for the first term in the right hand side of \eqref{secrate}, which represents the receiver DoF after jamming is applied.
 
With the eavesdropper completely blocked, it remains to show how the jamming signal directions are designed to achieve the maximum possible SDoF for the different regions stemming from relations between $(M, N_1, N_2,..., N_K, N_E)$.

\subsection{MIMO wiretap channel}
\subsubsection{Achievability for $M \geq N_1+N_E$}

For this region, the transmitter sends $\vec{V}^J$ in the null space of the receiver channel.

\begin{eqnarray}\label{partone}
\vec{V}^J=\text{Null} (\vec{H}_1) 
\end{eqnarray}
\\
This leaves the receiver with $N_1$ SDoF to decode and the upperbound is achieved.\\

\subsubsection{Achievability for $ N_1\leq M<N_1+N_E$}

For this region, the transmitter sends a jamming signal using precoder $\boldsymbol{V}^J$ composed of two parts to block the eavesdropper,
\begin{equation}\label{parts}
\vec{V}^J=\begin{array}{cc} [\vec{V}^J_1 & \vec{V}^J_2] \end{array}.
\end{equation} 

The first part $\vec{V}^J_1$ is sent in the null space of the receiver channel, as in~\eqref{partone}, with size $M\times J_1$, where $J_1=M-N_1$, while the second part, $\vec{V}^J_2$, is chosen randomly with size $M \times J_2$, where $J_2=N_E - J_1$. Consequently, the transmitter has $M-N_E$ available transmit directions, while the receiver has at least $M-N_E$ jamming--free receive directions to decode the $M-N_E$ securely transmitted streams and the upperbound is achieved.

\subsubsection{ Achievability for $M< N_1$}
For this region, the transmitter chooses $\vec{V}^J$ randomly, while this would jam some of the receiver space, this wont decrease the DoF because it is constrained by the transmit antennas\footnote{This is due to the fact that in this region, $M-N_E \leq N_1 - N_E$.}.
The receiver zero forces the jamming signal using the post-processing matrix $\vec{U}$ as in (\ref{zero}). Accordingly, $M-N_E$ SDoF can be sent and decoded by the receiver.\\

\begin{equation}\label{zero}
\vec{U}= [\vec{I} - \vec{a}\vec{a}^{-1}],
\end{equation}
 where
\begin{equation}
\vec{a}= \vec{H} \vec{V}^J.
\end{equation}

\subsection{The Two User Broadcast Channel}
\subsubsection{Achievability for  broadcast channel: $M\geq N_1 \geq N_2$}
For this region, the transmitter and both receivers agree on dedicated space for decoding at each receiver with sizes $d_1$ and $d_2$ for receivers one and two, respectively. The transmitter sends a jamming signal using precoder $\vec{V}^J$ in the null space of the union of the two dedicated decoding spaces. Let $\mathcal{A}_1$ and $\mathcal{A}_2$ be the decoding spaces of receiver one and two, respectively. Let $\vec{U}_1$ and $\vec{U}_2$ be the post processing matrices that project the received signal into $\mathcal{A}_1$ and $\mathcal{A}_2$, respectively.
\begin{equation}\label{comb}
\vec{V}^J=\text{Null}([\vec{U}_1\vec{H}_1 \text{; } \vec{U}_2\vec{H}_2])
\end{equation} 

\begin{proposition}
For this scheme the following region is achievable 
\begin{eqnarray}\label{bcregion}
\nonumber d^{bc}_1 \leq \min (N_1, M-N_E ) \\
\nonumber d^{bc}_2 \leq \min (N_2, M-N_E ) \\
d^{bc}_1+d^{bc}_2\leq \min (N_1+N_2, M-N_E ) 
\end{eqnarray}
\end{proposition}

\begin{proof}
The size of the nullspace in~\eqref{comb} is $M-d^{bc}_1-d^{bc}_2$, however, for blocking the eavesdropper the nullspace size must be $N_E$.  Therefor,
\begin{eqnarray}\
M-d^{bc}_1+d^{bc}_2= N_E \;\;\;\; \Rightarrow \;\;\;\;\;
d^{bc}_1+d^{bc}_2 = M-N_E.
\end{eqnarray}
It is easy to see that if $N_1+N_2< M-N_E$, the nullspace of the the receivers' channels can accommodate a jamming signal with size $N_E$. Hence, $N_1+N_2$ SDoF is achieved.
\end{proof}
				
\subsubsection{Achievability for  broadcast channel: $N_2 < M \leq N_1$}	
For this region, the jamming signal is divided into two parts as in~\eqref{parts}. The first part is of size $J_1=M-N_2$ and is sent in the nullspace of the second receiver's channel, thus, generating interference of size $J_1$ at the first receiver. The second part, with size $J_2= N_E - J_1$, is sent in random direction, hence, generating interference at both receivers. Consequently, the achievable region is
\begin{eqnarray}\label{bcregion}
\nonumber d^{bc}_2 \leq \text{min} (N_2, M-N_E ) \\
d^{bc}_1+d^{bc}_2\leq  M-N_E 
\end{eqnarray}

\begin{figure}\label{wire}
  \begin{center}
\hspace{-4mm} \includegraphics[width=.43\textwidth]{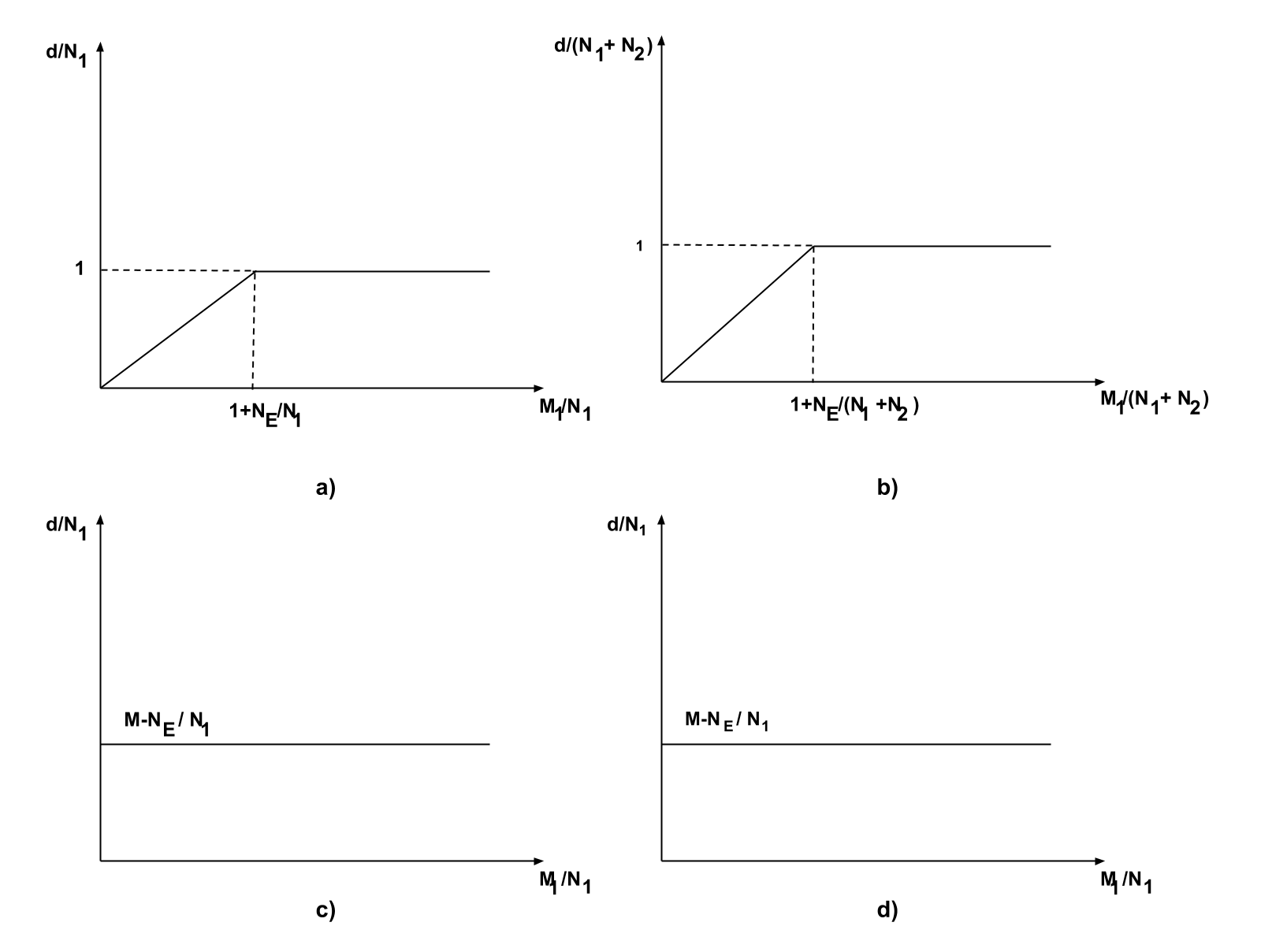}\vspace{-2mm}
\caption{ \small Sum SDof of: a) wiretap channel and b,c,d) two receiver broadcast channel, where for b) $M\geq N_1>N_2$, c) $N_1 \geq M > N_2$ and d) $N_1\geq N_2\geq M$/}
  \end{center} \vspace{-8mm}
	\normalsize
  \end{figure}
    
	
\subsubsection{Achievability for  broadcast channel: $ M < N_2 \leq N_1$}		
For this region, the jamming signal direction is totally randomly chosen, and the SDoF region is
\begin{eqnarray}\label{bcregion3}
d^{bc}_1+d^{bc}_2 \leq M-N_E 
\end{eqnarray}

\subsection{$K$--user broadcast channel}
Consider the case of $K$--user broadcast channel with $M$ transmit antennas and $N$ receive antennas at each receiver.

 \textit{Achievability}: For this system, the transmitter and the receivers agree on a decoding space, $\mathbb{A} = \bigcup_{i = 1}^{K} \mathcal{A}_i$, of size $\text{min}(M-N_E, KN)$, where $\mathcal{A}_i$ is the decoding space for receiver $i$. Let $\vec{U}_i$ be the $l_i \times N$ postprocessing matrix projecting the received signal into $\mathcal{A}_i$.  The jamming signal is transmitted in the nullspace of $\mathbb{A}$. Therefor,

\begin{eqnarray}
\vec{V}^J&=\text{Null}(\left[\vec{U}_1 \vec{H}_1\text{; }\vec{U}_2 \vec{H}_2 \cdots \vec{U}_K \vec{H}_K\right])
\end{eqnarray}
and $l_i$ represents the SDoF of receiver $i$ such that $\sum_{i=1}^K l_i=\text{min}(M-N_E, KN)$. Hence,
\begin{equation}
\sum_{i=1}^K d^{bc}_i = \begin{cases} M-N_E & M\leq KN + N_E,\\  KN & M > KN +N_E.\end{cases}  
\end{equation}

\section{Conclusion}
\label{con}

We studied the wiretap and the broadcast channel with multiple antennas at the transmitter, legitimate receivers and eavesdroppers in the existence of unknown eavesdroppers. A new upperbound was established and a new achievable DoF region was provided, the secure DoF regions were identified. We proved that transmitter sent jamming is SDoF optimal even if it has no eavesdroppers' CSI.

%
%
%
 %
%
%
\end{document}